\documentclass{llncs}
\usepackage{graphicx}
\usepackage{mathrsfs}
\usepackage{amssymb}
\usepackage{amsmath}
\input{psfig.sty}
\usepackage{latexsym}

\newcommand{\prj}               {\mathit{\,prj\,\,}}
\newcommand{\CHOP}              {\boldsymbol{;} }

\newcommand{\true}              {\mathit{true}}
\newcommand{\false}             {\mathit{false}}
\newcommand{\interp}            {{\cal{I}}}
\newcommand{\emptyy}            {\varepsilon}

\newcommand{\DEF}               {\stackrel{\rm def}{=}}
\newcommand{\NFG}               {\mathit{NFG}}

\newcommand{\SKIP}              {\mathit{skip}}
\newcommand{\Prj}               {\mathit{Prj}}
\newcommand{\fin}               {\mathit{fin}}
\newcommand{\halt}              {\mathit{halt}}
\newcommand{\keep}              {\mathit{keep}}
\newcommand{\more}              {\mathit{more}}

\newcommand{\len}               {\mathit{len}}
\newcommand{\Frame}             {\mathit{frame}}

\newtheorem{Expl}{Example}
\newtheorem{Def}{Definition}
\newtheorem{Thm}{Theorem}

\newtheorem{Cor}[Thm]{Corollary}

\def\squareforqed{\hbox{\rlap{$\sqcap$}$\sqcup$}}
\def\qed{\ifmmode\squareforqed\else{\unskip\nobreak\hfil
\penalty50\hskip1em\null\nobreak\hfil\squareforqed
\parfillskip=0pt\finalhyphendemerits=0\endgraf}\fi}

\newcounter{statement}
\def\stmnum{\hbox to .01pt{}\rlap{\rm \hskip -\displaywidth\thestatement.}}
\def\stm{\refstepcounter{statement}\topsep 2pt \trivlist \item[]\leavevmode
\hbox to\linewidth\bgroup $ \displaystyle \hskip\leftmargini}
\def\endstm{$\hfil \displaywidth\linewidth\stmnum\egroup \endtrivlist}

\newenvironment{Proof}{%
\begin{list}{}{\setlength{\topsep}{\jot}\setlength{\parsep}{\topsep}%
\addtolength{\parsep}{-0.3\parsep}\setlength{\leftmargin}{0pt}}%
\parindent 4ex
\item[]\setcounter{statement}{0}\textbf{Proof:}}{\end{list}}

       \title{Probabilistic Model Checking on Propositional Projection Temporal Logic
        \thanks{This research is supported by the NSFC No.60721061, China Postdoctoral Science Foundation and the CAS Innovation Programs.}}
       \author{Xiaoxiao Yang }
      \institute{  State Key Laboratory of Computer Science
                 \\ Institute of Software, Chinese Academy of Sciences
                 \\ Beijing, 100190, China\\ xxyang@ios.ac.cn}

\begin{document}
          \maketitle

          \begin{abstract}
             Propositional Projection Temporal Logic (PPTL) is a useful formalism for
             reasoning about period of time in hardware and software systems and
             can handle both sequential and parallel compositions.
             In this paper, based on discrete time Markov chains, we investigate the probabilistic model checking approach for PPTL towards verifying
             arbitrary linear-time properties.
             We first define a normal form graph, denoted by $\NFG_{\mathit{inf}}$,
             to capture the infinite paths of PPTL formulas.
             Then we present an algorithm to generate the $\NFG_{\mathit{inf}}$.
             Since discrete-time Markov chains are the deterministic probabilistic models, we further give
             an algorithm to determinize  and minimize the nondeterministic  $\NFG_{\mathit{inf}}$ following the Safra's construction.
          \\[1em]
          \textbf{Keywords:} projection temporal logic, probabilistic model checking, Markov chains, normal form graph.
          \end{abstract}

           \section{Introduction}
           Traditional model checking techniques focus on a systematic check of the validity of
           a temporal logic formula on a precise mathematical model.  The answer to the
           model checking question is either
           true or false. Although this classic approach is
           enough to specify and verify boolean temporal properties,  it does
           not allow to reason about stochastic nature of systems. In real-life systems, there are many phenomena
           that can only be modeled by considering their stochastic characteristics. For this purpose,
           probabilistic model checking is proposed as a formal verification technique for the analysis of
           stochastic systems. In order to model random phenomena, discrete-time Markov chains,
           continuous-time Markov chains and Markov decision processes are widely used in probabilistic model checking.

           Linear-time property is a set of infinite paths. We can use linear-time temporal logic (LTL)
           to express $\omega$-regular properties. Given a finite Markov chain $M$ and an $\omega$-regular property $Q$,
           the probabilistic model checking problem for LTL is
           to compute the probability of accepting runs
           in the product Markov chain $M$ and a deterministic Rabin automata (DRA) for $\neg Q$ \cite{Katoen}.

           Among linear-time temporal logics, there exists a number of \emph{choppy logics} that are based on chop ($\CHOP$) operators.
           Interval Temporal Logic (ITL) \cite{Mos83} is one kind of choppy
           logics, in which
           temporal operators such as \emph{chop}, \emph{next} and \emph{projection} are defined. Within the ITL developments,
           Duan, Koutny and Holt, by introducing a new projection construct  $(p_1,\ldots,p_m) \prj q$,
           generalize ITL to infinite time intervals. The new interval-based temporal logic is called
           Projection Temporal Logic (PTL) \cite{ZCL07}. PTL is a useful formalism for reasoning about period of time for
           hardware and software systems. It can handle both sequential and parallel compositions, and offer useful and practical proof
           techniques for verifying concurrent systems
           \cite{ZN08,ZCL07}. Compared with LTL,  PTL can describe more linear-time properties.
            In this paper, we investigate the probabilistic model checking on Propositional PTL (PPTL).

           There are a number of reasons for being interested in
           projection temporal logic language. One is that
           projection temporal logic can express various imperative programming constructs (e.g. while-loop) and has executable subset \cite{DYK08,YD08}.
           In addition, the expressiveness of projection temporal
           logic is more powerful than the classic point-based
           temporal logics such as LTL   since the temporal logics
           with \emph{chop star} ($*$) and \emph{projection} operators are equivalent to
            $\omega$-regular languages, but LTL cannot express all $\omega$-regular properties \cite{wolper}.
           Furthermore,
           the key construct used in PTL is
           the new projection operator $(p_1,\ldots, p_m)~\prj~q$
           that can be thought of as a combination of the parallel
           and the projection operators in ITL.
           By means of the projection construct, one can define fine- and
           coarse-grained concurrent behaviors in a flexible and readable
           way. In particular, the sequence of
           processes $p_1,\ldots, p_m$ and process $q$ may
           terminate at different time points.

          In the previous work \cite{DYK08,YD08,ZCL07}, we have presented a \emph{normal form} for any PPTL formula.
          Based on the normal form, we can construct a semantically equivalent graph, called \emph{normal form graph} (NFG).
          An infinite (finite) interval that satisfies a PPTL formula will correspond to an infinite (finite) path in NFG.
          Different from Buchi automata, NFG is exactly the model of a PPTL formula. For any unsatisfiable PPTL formula, NFG
          will be reduced to a false node at the end of the construction.  NFG consists of both finite and
          infinite paths. But for concurrent stochastic systems, here we
          only consider infinite cases. Therefore, we define $\NFG_{\mathit{inf}}$ to denote an NFG only with infinite paths.
          To capture the accurate semantics for PPTL formulas with infinite intervals,
          we adopt Rabin acceptance condition as accepting states in $\NFG_{\mathit{inf}}$. In addition, since  Markov chain $M$ is a
          deterministic probabilistic model, in order to guarantee that the product of $M\otimes \NFG_{\mathit{inf}}$ is also a Markov chain,
          we give an algorithm for deterministic $\NFG_{\mathit{inf}}$, in the spirit of Safra's construction for deterministic
          Buchi-automata.

\begin{figure}
            \includegraphics[width=9cm]{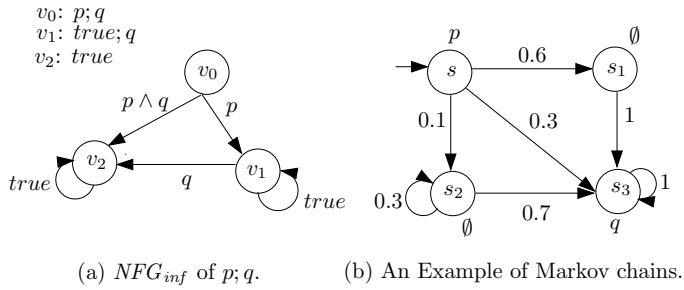}\\
            \caption{A Simple Example for Probabilistic Model Checking on PPTL.}\label{eg-chop}
          \end{figure}

          To make this idea clear, we now consider a simple
          example shown in Figure \ref{eg-chop}. The definitions of NFGs and Markov chains are
          formalized in the subsequent sections. Let $p ~\CHOP~ q$ be a \emph{chop} formula in PPTL, where $p$ and
          $q$ are atomic propositions.
          $\NFG_{\mathit{inf}}$ of $p~\CHOP~q$ is constructed in Figure \ref{eg-chop}(a), where
          nodes $v_0$, $v_1$ and $v_2$ are temporal formulas, and edges are state
          formulas (without temporal operators). $v_0$ is an initial
          node. $v_2$ is an acceptance node recurring for
          infinitely many times, whereas $v_1$ appears finitely
          many times. Figure \ref{eg-chop}(b) presents a Markov
          chain with initial state $s$. Let path $\mathit{path} = \langle s, s_1, s_3
          \rangle$. We can see that $\mathit{path}$ satisfies $p ~\CHOP
          ~q$ with probability 0.6. Based on the product of Markov
          chain and $\NFG_{\mathit{inf}}$, we can compute the whole probability
          that the Markov chain satisfies $p~\CHOP~q$.

           Compared with Buchi automata,  NFGs have the following advantages that are more suitable for verification for interval-based temporal logics.
           \\
           (i) NFGs are beneficial for unified verification approaches based on the same formal notation.
           NFGs can not only be regarded as  models of specification language PTL,
            but also as models of Modeling Simulation and Verification Language (MSVL)\cite{DYK08,YD08},
            which is an executable subset of PTL.
            Thus, programs and their properties can be written in the same language, which avoids the transformation between different notations.
           \\
           (ii) NFGs can accept both finite words and infinite
           words.  But Buchi automata can only accept infinite words. 
           Further, temporal operators \emph{chop} ($p ~\CHOP ~q$), \emph{chop star}
          ($p^*$), and \emph{projection} can be readily transformed to NFGs. \\
           (iii) NFGs and PPTL formulas are semantically equivalent. That is, every path in NFGs corresponds to a model of PPTL formula.
            If some formula is false, then its NFG will be a false node.
            Thus, satisfiability in PPTL formulas can be
            reduced to NFGs construction. But for any LTL formula, the satisfiability problem
            needs to check the emptiness problem of Buchi automata.

            The paper is organized as follows. Section 2 introduces PPTL briefly. Section 3 presents the (discrete time) Markov chains.
            In Section 4, the probabilistic model checking approach for PPTL is investigated.
            Finally, conclusions are drawn in Section 5.


           \section{Propositional Projection Temporal Logic}

            The underlying logic we use is Propositional Projection Temporal Logic (PPTL).
            It is a variation of Propositional Interval Temporal Logic (PITL).
            \begin{Def}\rm
           Let $AP$ be a finite set of atomic propositions. PPTL formulas over $AP$ can be defined as follows:

           \[
              Q ::= \pi \mid \neg Q \mid \bigcirc Q \mid  Q_1 \wedge Q_2  \mid (Q_1, \ldots, Q_m) \prj Q \mid Q^+
           \]
           where $\pi \in AP$, $Q, Q_1,\ldots, Q_n$ are PPTL formulas, $\bigcirc$ (next), $\prj$ (projection) and $+$ (plus) are basic temporal operators.
           \end{Def}

           A formula is called a \emph{state} formula if it does
           not contain any temporal operators, i.e.,  \emph{next} ($\bigcirc$),
           \emph{projection} ($\prj$) and \emph{chop-plus} (${}^+$); otherwise
           it is a \emph{temporal} formula.

           An interval $\sigma = \langle s_0,s_1,\ldots\rangle$ is a non-empty sequence of states,
           where $s_i~(i \geq 0)$ is a state mapping from $AP$ to $B= \{true, false\}$.
           The length, $|\sigma|$, of $\sigma$ is $\omega$ if $\sigma$ is infinite, and the number of states minus 1
           if $\sigma$ is finite.  To have a uniform notation for both finite and infinite intervals, we
           will use \emph{extended integers} as indices. That is, for set $N_0$ of non-negative integer and $\omega$,
           we define $ N_{\omega} = N_0 \cup \{\omega\}$,
           and extend the comparison operators: $=, <, \leq,$ to
           $N_{\omega}$ by considering $\omega = \omega$, and for all $i \in N_0, i < \omega$.
           Moreover,  we define $\preceq$ as $\leq-\{(\omega,\omega)\}$.

        To define the semantics of the projection construct we need an auxiliary operator. Let $\sigma = \langle s_0, s_1,...\rangle$ be an
        interval and $r_1,\ldots,r_h$ be integers ($h\geq 1$) such that $0\leq r_1\leq \ldots \leq r_h\preceq |\sigma| $.
       \[
         \sigma \downarrow (r_1,\ldots,r_h)
         \DEF
         \langle s_{t_1}, s_{t_2},\ldots, s_{t_l}\rangle
        \]

\begin{figure}[htbp]
             \flushleft\includegraphics[width=10cm]{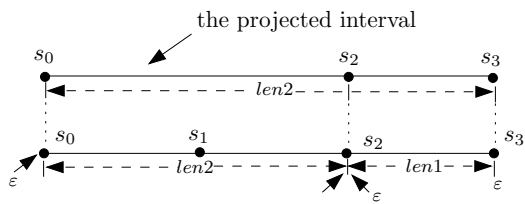}\\
             \caption{A projected interval.}\label{projection}
             \end{figure}

        The \emph{projection} of $\sigma$ onto $r_1,\ldots,r_h$ is the interval (called projected interval) where  $t_1,\ldots,t_l$ are
        obtained from $r_1,\ldots,r_h$ by deleting all duplicates. In other words, $t_1,\ldots,t_l$ is the longest strictly increasing
        subsequence of $r_1,\ldots,r_h$. For example,
        $
               \langle s_0, s_1, s_2, s_3 \rangle \downarrow (0,2,2,2,3)
              =
               \langle s_0, s_2, s_3 \rangle.
        $ As depicted in Figure \ref{projection}, the projected interval $\langle s_0, s_2, s_3\rangle$ can be obtained by using $\downarrow$ operator to
        take the endpoints of each process $\emptyy, len(2), \emptyy, \emptyy,
        len(1)$.

           An interpretation for a PPTL formula is a tuple $\interp =(\sigma,i,k,j)$, where $\sigma$ is an interval,
           $i,k$ are integers, and $j$ an integer or $\omega$ such that $i \leq k \preceq j$. Intuitively, $(\sigma,i,k,j)$
           means that a formula is interpreted over a subinterval
           $\sigma_{(i,..,j)}$ with the current state being $s_k$. The satisfaction relation ($\models$) between interpretation $\interp$ and formula
           $Q$ is inductively defined as follows.
          \begin{center}
          \begin{enumerate}
                \item $\interp \models \pi$ iff $s_k[\pi]= true$
                \item $\interp \models \neg Q$ iff $\interp \nvDash Q$
                \item $\interp \models Q_1 \wedge Q_2$ iff $\interp \models
                Q_1$ and $\interp \models Q_2$
                \item $\interp \models \bigcirc Q$ iff $k < j$ and $(\sigma, i, k+1, j) \models Q$
                \item $\interp \models (Q_1,\ldots,Q_m) \prj Q$  iff there are~
                $k=r_0\leq r_1\leq\ldots \leq r_m \preceq j$~\mbox{such that }
                $(\sigma,i,r_0,r_1)\models Q_1 ~\mbox{and}$
                $(\sigma,r_{l-1},r_{l-1}, r_l)\models Q_l~\mbox{for all}~1<l\leq m ~\mbox{and}$
                $ (\sigma',0,0,|\sigma'|)$ $\models Q ~\mbox{for}~ \sigma'~\mbox{given  by}:$
                \\
                $ (a) ~r_m<j ~\mbox{and}~ \sigma' = \sigma \downarrow (r_0,\ldots,r_m)~{\cdot}~\sigma_{(r_m+1,..,j)}$
                \\
                $(b) ~ r_m=j ~ \mbox{and}~ \sigma'=\sigma \downarrow (r_0,\ldots,r_h) ~\mbox{for some}~ 0\leq h\leq  m$.
                \item $\interp \models Q^+$
                $\mbox{iff there are finitely many } r_0,\ldots,r_n \mbox{ and }$
                $k= r_0\leq r_1 \leq \ldots \leq r_{n-1}\preceq r_n =j ~(n\geq 1)$\\
                $\mbox{such that }$ $(\sigma, i, r_0, r_1)\models Q  \mbox{ and }$
                $(\sigma,r_{l-1}, r_{l-1}, r_l)\models Q~\mbox{for all}~1<l\leq n$ or\\
                $j= \omega$ and there are infinitely many integers $k= r_0 \leq r_1 \leq r_2 \leq \ldots$ such that $\lim\limits_{i\rightarrow\infty}
                r_i = \omega$ and $(\sigma, i, r_0, r_1) \models Q$ and for $l >1,
                (\sigma, r_{l-1}, r_{l-1}, r_l) \models Q$.
        \end{enumerate}
        \end{center}

          A PPTL formula $Q$ is satisfied by an interval $\sigma$, denoted
          by $\sigma \models Q$, if $(\sigma, 0, 0, |\sigma|) \models
          Q$. A formula $Q$ is called satisfiable, if $\sigma \models
          Q$. A formula $Q$ is valid, denoted by $\models Q$, if
          $\sigma \models Q$ for all $\sigma$.  Sometimes, we denote $\models p\leftrightarrow q$ (resp.
$\models p\rightarrow q$) by $p\approx q$ (resp.$\hookrightarrow$ )
and $\models \Box(p \leftrightarrow q)$ (resp. $\models \Box(p
\rightarrow q)$) by $p \equiv q$ (resp. $p\supset q$), The former is
called \emph{weak equivalence (resp. weak implication)} and the
latter \emph{strong equivalence (resp. strong implication)}.

Figure~\ref{fig-formulas} below shows us some useful formulas
derived from elementary PTL formulas. $\emptyy$ represents the final
state and $\more$ specifies that the current state is a non-final
state; $\Diamond P$ (namely \emph{sometimes} $P$) means that $P$
holds eventually in the future including the current state;
 $\Box P$ (namely \emph{always} $P$) represents that $P$ holds always
in the future from now on; $\bigodot P$ (\emph{weak next}) tells us
that either the current state is the final one or $P$ holds at the
next state of the present interval; $\Prj (P_1,\ldots,P_m)$
represents a \emph{sequential} computation of $P_1, \ldots, P_m$
since the projected interval is a singleton; and $P\;\CHOP \;Q$ ($P$
\emph{chop} $Q$) represents a computation of $P$ followed by $Q$,
and the intervals for $P$ and $Q$ share a common state. That is, $P$
holds from now until some point in future and from that time point
$Q$ holds. Note that $P \;\CHOP \;Q$ is a strong chop which always
requires that $P$ be true on some finite subinterval. $len(n)$
specifies the distance $n$ from the current state to the final state
of an interval; $\SKIP$ means that the length of the interval is one
unit of time. $\fin(P)$ is $\true$ as long as $P$ is $\true$ at the
final state while $\keep(P)$ is $\true$ if $P$ is true at every
state but the final one. The formula $\halt(P)$ holds if and only if
formula $P$ is $\true$ at the final state.

\begin{figure}[htpb]
\[
\begin{array}{lcl}
      \emptyy
    & \DEF
    & \neg\bigcirc\true
    \\
      len(n)
    & \DEF
    & \left\{
      \begin{array}{ll}
      \emptyy
      & \mbox{if} ~n=0
      \\
      \bigcirc len(n-1)
      & \mbox{if} ~n>1
      \end{array}
      \right.
\\
    \Box P
    & \DEF
    & \neg\Diamond\neg P
    \\
    \SKIP
    & \DEF
    & len(1)
\\
      \Prj (P_1,\ldots,P_m)
    & \DEF
    & (P_1,\ldots,P_m) \prj \emptyy
    \\ \fin(P)
    & \DEF
    & \Box(\emptyy \rightarrow P)
\\
      P\;\CHOP \;Q
    & \DEF
    & \Prj \;(P,Q)
   \\ \keep(P)
    & \DEF
    & \Box(\neg \emptyy \rightarrow P)
\\
      \more
    & \DEF
    & \neg\emptyy
    \\ \halt(P)
    & \DEF
    & \Box(\emptyy \leftrightarrow P)
\\
      \Diamond P
    & \DEF
    & \Prj(\true, P)
    \\  \bigodot P
    & \DEF
    & \emptyy  \vee \bigcirc P
\end{array}
\]
\caption{Derived PPTL formulas.} \label{fig-formulas}
\end{figure}

\subsubsection*{An Application of Projection Construct}
\begin{Expl}\rm
We present a simple application of projection construct about a pulse generator for variable $x$ which can assume two values: 0 (low) and 1 (high).

We first define two types of processes: The first one is $hold(i)$ which is executed over an interval of length $i$ and ensures that the value of $x$ remains constant in all but the final state,
\[
      hold(i) \DEF \Frame(i) \wedge \len(i)
\]

The other is $switch(j)$ which is ensures that the value of $x$ is first set to 0 and then changed at every subsequent state,
\[
      switch(j) \DEF x=0 \wedge \len(j) \wedge \Box(\more \rightarrow \bigcirc x= 1-x)
\]

Having defined $hold(i)$ and $switch(j)$, we can define the pulse generators with varying numbers and length of low and high intervals for $x$,
\[
     pulse(i_1, \ldots, i_k) \DEF  (hold(i_1), \ldots, hold(i_k)) \prj switch(k)
\]

 For instance, a pulse generator
 \[
   \begin{array}{lrl}
   pulse(3, 5, 3, 4) &\DEF& (hold(3), hold(5), hold(3), hold(4))
   \prj switch(4)
   \end{array}
 \] can be shown in Figure \ref{projection-example}.

\begin{figure}[t!]
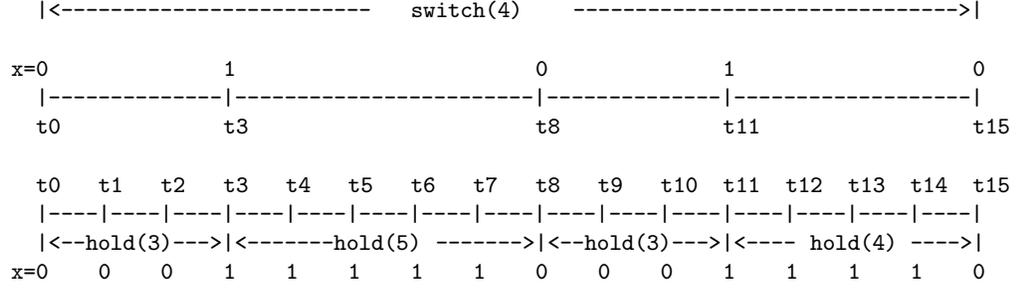

\begin{center}
{\small
\begin{verbatim}
     |<-------------------------   switch(4)    ------------------------------->|

   x=0              1                        0              1                   0
     |--------------|------------------------|--------------|-------------------|
     t0             t3                       t8             t11                 t15

     t0   t1   t2   t3   t4   t5   t6   t7   t8   t9   t10  t11  t12  t13  t14  t15
     |----|----|----|----|----|----|----|----|----|----|----|----|----|----|----|
     |<--hold(3)--->|<-------hold(5) ------->|<--hold(3)--->|<---- hold(4) ---->|
   x=0    0    0    1    1    1    1    1    0    0    0    1    1    1    1    0

  \end{verbatim}
 } \caption{A Pulse Generator}\label{projection-example}
 \end{center}
\end{figure}
\end{Expl}

            Let $Q$ be a PPTL formula and $Q_p \in AP$ be a set of atomic propositions in $Q$.
            Normal form of PPTL formulas can be defined as follows.

          \begin{Def}\rm
          A PPTL formula $Q$ is in \emph{normal form} if
          \[
             Q \equiv (\bigvee\limits_{j=0}^{n_0} Q_{e_j} \wedge \emptyy) \vee ( \bigvee\limits_{i=0}^{n} Q_{c_i} \wedge \bigcirc Q_{f_i})
          \]
          where $Q_{e_j} \equiv \bigwedge\limits_{k=1}^ {m_0} \dot{q_{jk}}, Q_{c_i} \equiv \bigwedge\limits_{h=1}^ m \dot{q_{ih}}$,
           $|Q_p|=l$, $1 \leq m_0 \leq l$, $1 \leq m \leq l$;
          $q_{jk}, q_{ih} \in
          Q_p$, for any $r \in Q_p$, $\dot{r}$ means $r$ or $\neg
          r$; $Q_{fi}$ is a general PPTL formula. For convenience, we
          often write $Q_e \wedge \emptyy$ instead of $\bigvee\limits_{j=0}^{n_0} Q_{e_j} \wedge \emptyy$
          and $\bigvee\limits_{i=0}^{n} Q_{i} \wedge \bigcirc
          Q_{i}'$ instead of $\bigvee\limits_{i=0}^{n} Q_{c_i} \wedge \bigcirc
          Q_{f_i}$. Thus,
          \[
             Q \equiv (Q_e \wedge \emptyy) \vee (\bigvee\limits_{i=0}^{n} Q_{i} \wedge \bigcirc
          Q_{i}')
          \]
          where $Q_e$ and $Q_i$ are state formulas.
          \end{Def}

          \begin{Thm}\rm
           For any PPTL formula $Q$, there is a normal form $Q'$  such that $Q \equiv Q'$. \cite{ZCL07}
          \end{Thm}

         \section{Probabilistic System}
         We model probabilistic system by \emph{(discrete-time)} \emph{Markov chains} (DTMC).
         Without loss of generality, we assume that a DTMC has a unique initial state.

         \begin{Def}\rm
         A Markov chain is a tuple
         $M=(S, \mathit{Prob}, \iota_{init}, AP, L)$,
         where $S$ is a countable, nonempty set of states; $\mathit{Prob}: S \times S\rightarrow [0,1]$ is the transition probability function such that
         $\sum\limits_{s' \in S} \mathit{Prob}(s, s') =1$; $ \iota_{init}: S \rightarrow [0,1]$ is the initial distribution such that
         $\sum\limits_{s\in S}\iota_{init}(s) =1$, and  $AP$ is a set of atomic propositions and $L: S \rightarrow 2^{AP}$ a labeling function.
         \end{Def}

         As in the standard theory of Markov processes \cite{KS60},
         we need to formalize a probability space of $M$ that can be defined as $\psi_M =(\Omega, \mathit{Cyl}, Pr)$, where $\Omega$ denotes the set of all infinite
         sequences of states $\langle s_0, s_1, \ldots \rangle$  such that $\mathit{Prob} (s_i, s_{i+1}) >0$ for all $i \leq 0$, $\mathit{Cyl}$
         is a $\sigma$-algebra generated by the \emph{basic cylindric sets}:
         \[
         \mathit{Cyl}(s_0, \ldots, s_n) = \{ path \in \Omega \mid path = s_0, s_1, \ldots, s_n, \ldots\}
         \]
         and $Pr$ is a probability distribution defined by

         \begin{eqnarray*}
             Pr^M (\mathit{Cyl}(s_0, \ldots, s_n))& = &  \mathit{Prob} (s_0, \ldots, s_n)
                                            \\& =&  \prod\limits_{0 \leq i < n} \mathit{Prob} (s_i, s_{i+1})
         \end{eqnarray*}

         If $p$ is a path in DTMC $M$ and $Q$ a PPTL formula, we often write $p \models Q$ to mean that a path in
         DTMC satisfies the given formula $Q$. Let $\mathit{path(s)}$ be a set of paths in DTMC starting with state $s$. The
         probability for $Q$ to hold in state $s$ is denoted by $Pr^M(s \models Q)$, where
         $Pr^M(s \models Q) = Pr^M_s \{p \in \mathit{path(s)} \mid p \models Q\}$.

          \section{Probabilistic Model Checking for PPTL}
          In \cite{ZCL07}, it is shown that any PPTL formulas can be rewritten
          into normal form, where a graphic description for normal form called  Normal Form
          Graph (NFG) is presented. NFG is an important basis of decision procedure
          for satisfiability and model checking for PPTL. In this
          paper, the work reported  depends on the NFG to investigate the probabilistic
          model checking  for PPTL.

          However, there are
          some differences on NFG between our work and the previous work
          in \cite{DYK08,YD08,ZCL07}. First, NFG consists of finite paths and
          infinite paths. For concurrent stochastic systems,  we
          only consider to verify $\omega$-regular properties. Thus, we
          are supposed to concern with all the infinite paths of NFG. These infinite paths are denoted by
          $\NFG_{\mathit{inf}}$.
          Further,  to define the nodes which recur for finitely many times, \cite{ZCL07} uses Labeled NFG (LNFG)  to
          tag all
          the nodes in finite cycles with $F$. But it  can not identify all the possible acceptance cases. As  the
          standard acceptance conditions in  $\omega$-automata,  we adopt Rabin acceptance condition to precisely define the
          infinite paths in $\NFG_{\mathit{inf}}$. In addition, since  Markov chain $M$ is a
           deterministic probabilistic model, in order to guarantee that the product of $M\otimes \NFG_{\mathit{inf}}$ is also a Markov chain,
           the $\NFG_{\mathit{inf}}$ needs to be deterministic. 
           Thus, following the Safra's construction for deterministic
           automata, we design an algorithm to  obtain a deterministic
           $\NFG_{\mathit{inf}}$.

          \subsection{Normal Form Graph}
          In the following, we first give a general definition of NFG for PPTL
          formulas.
          \begin{Def}[Normal Form Graph \cite{DYK08,ZCL07}]\rm
          For a PPTL formula $P$, the set $V(P)$ of nodes and the set of $E(P)$ of edges connecting nodes in $V(P)$ are
          inductively defined as follows.
          \begin{enumerate}
          \item $P \in V(P)$;
          \item For all $Q \in V(P) / \{\emptyy, \false\}$, if $Q \equiv (Q_e \wedge \emptyy) \vee (\bigvee\limits_{i=0}^{n} Q_{i} \wedge \bigcirc
          Q_{i}')$, then $\emptyy \in V(P)$, $(Q, Q_e, \emptyy) \in E(P)$; $Q_i' \in V(P)$, $(Q, Q_i, Q_i') \in E(P)$ for all $i$, $1 \leq i \leq n$.
          \end{enumerate}
          The NFG of PPTL formula $P$ is the directed graph $G = (V(P), E(P))$.
          \end{Def}

           A finite path for formula $Q$ in NFG
           is a sequence of nodes and edges from the root
          to node $\emptyy$. while an infinite path
          is an infinite sequence of nodes and edges
          originating from the root.

          \begin{Thm}[Finiteness of NFG]\label{finiteness}\rm
          For any PPTL formula $P$, $|V(P)|$ is finite \cite{ZCL07}.
          \end{Thm}

          Theorem \ref{finiteness} assures that the number of nodes in NFG is finite.
          Thus, each satisfiable formula of PPTL is satisfiable by a finite transition system (i.e., finite NFG).
          Further, by the finite model property,  the satisfiability of PPTL is decidable.
          In \cite{ZCL07}, Duan \emph{etal} have given a decision procedure for PPTL formulas based on NFG.

          To verify $\omega$-regular properties, we need to consider the infinite paths
          in NFG. By ignoring all the finite paths, we can obtain a subgraph only with infinite paths,
          denoted  $\NFG_{\mathit{inf}}$.

          \begin{Def}\label{NFG-inf}\rm
          For a PPTL formula $P$, the set $V_{\mathit{inf}}(P)$ of nodes and the set of $E_{\mathit{inf}}(P)$ of edges
          connecting nodes in $V_{\mathit{inf}}(P)$ are
          inductively defined as follows.

          \begin{enumerate}
          \item $P \in V_{\mathit{inf}}(P)$;
          \item For all $Q \in V_{\mathit{inf}}(P)$, if $Q \equiv (Q_e \wedge \emptyy) \vee (\bigvee\limits_{i=0}^{n} Q_{i} \wedge \bigcirc
          Q_{i}')$, then $Q_i' \in V_{\mathit{inf}}(P)$, $(Q, Q_i, Q_i') \in E_{\mathit{inf}}(P)$ for all $i$, $1 \leq i \leq n$.
          \end{enumerate}

          Thus, $\NFG_{\mathit{inf}}$ is a directed graph $G' = (V_{\mathit{inf}}(P), E_{\mathit{inf}}(P))$.
          Precisely, $G'$ is a subgraph of $G$ by deleting all the finite path from node $P$ to node $\emptyy$.
          \end{Def}

          In fact, a finite path in the NFG of a formula $Q$ corresponds
          to a model (i.e., interval) of $Q$. However, the result does not hold
          for the infinite case since not all of the infinite paths in NFG can
          be the models of $Q$. Note that, in an infinite path, there must
          exist some nodes which appear infinitely many times,
          but there may have other nodes that can just recur for finitely many
          times. To capture the precise semantics model of formula
          $Q$, we make use of Rabin acceptance condition as the
          constraints for nodes that must recur finitely.

          \begin{Def}\rm
          For a PPTL formula $P$, $\NFG_{\mathit{inf}}$ with Rabin acceptance condition is  defined as
          $G_{Rabin} = (V_{\mathit{inf}}(P), E_{\mathit{inf}}(P), v_0, \Omega)$,
          where $V(P)$ is the set of nodes and $E(P)$ is the set of directed edges between $V(P)$, $v_0 \in V(P)$ is the initial node,
          and $\Omega=\{(E_1, F_1), \ldots, (E_k, F_k)\}$ with $E_i, F_i \in V(P)$ is Rabin acceptance condition.
          We say that: an infinite path is a model of the formula $P$ if there exists an infinite run $\rho$ on the path such that
          \[
            \exists (E, F) \in \Omega. (\rho \cap E = \emptyset) \wedge (\rho \cap F \neq \emptyset)
          \]
          \end{Def}

          \begin{Expl}\rm
          Let $Q$ be PPTL formulas. The normal form of $\Diamond Q$ are as follows.
          \begin{eqnarray*}
          \Diamond Q & \equiv & \true \;\CHOP\; Q\\
                     & \equiv & (\emptyy \vee \bigcirc \true) \; \CHOP \; Q \\
                     & \equiv & (\emptyy \; \CHOP \; Q) \vee (\bigcirc \true \; \CHOP \; Q)\\
                     & \equiv & Q \vee \bigcirc (\true \CHOP Q) \\
                     & \equiv & (Q \wedge \emptyy) \vee (Q \wedge \bigcirc \true) \vee \bigcirc \Diamond Q\\
                     & \equiv & (Q \wedge \emptyy) \vee (Q \wedge \bigcirc (\emptyy \vee \bigcirc \true)) \vee \bigcirc \Diamond Q
          \end{eqnarray*}

          The NFG and $\NFG_{\mathit{inf}}$ with Rabin
          acceptance condition of $\Diamond Q$ are depicted in Figure \ref{NFG-expl}.
          By the semantics of formula $\Diamond Q$ (see Figure
          \ref{fig-formulas}), that is, formula $Q$ holds eventually in the
          future including the current state, we can know that node
          $\Diamond Q$ must cycle for finitely many times and node $T$ (i.e., $\true$) for infinitely many times.

          \begin{figure*}[htpb]
          \centering\includegraphics[width=15cm]{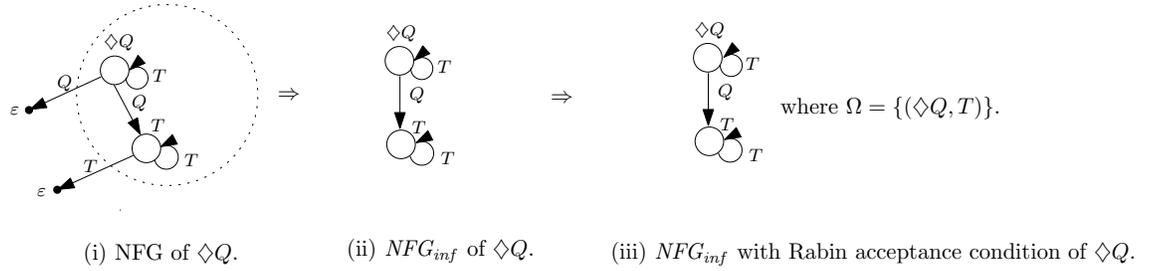} \caption{NFG of $\Diamond Q$.} \label{NFG-expl}
          \end{figure*}
          \end{Expl}

           \begin{table*}[thpb]
          \centering \caption {Algorithm for constructing $\NFG_{\mathit{inf}}$ with Rabin condition for a PPTL formula.} \label{Alg-NFG-INF}
          \begin{tabular}{|l|}
          \hline\\[-.7em]
          \textbf{Function} $\NFG_{\mathit{inf}}(Q)$\\[.2em]
          /*precondition: Q is a PPTL formula, NF(Q) is the normal form for $Q$ */ \\[.2em]
          /*postcondition: $\NFG_{\mathit{inf}}(Q)$ outputs $\NFG_{\mathit{inf}}$ with Rabin condition of $Q$,
          \\[.2em]~~~~~~~~~~~~~~~~~~~~~~
          $G_{\mathit{Rabin}} = (V_{\mathit{inf}}(Q), E_{\mathit{inf}}(Q), v_0, \Omega)$ */ \\[.2em]
          \hline\\[-.7em]
          \textbf{begin function}\\[.2em]
          $~~V_{\mathit{inf}} (Q) =\{Q\}; E_{\mathit{inf}}(Q) =\emptyset; \textbf{visit}(Q)=0;
          v_0 = Q; E=F=\emptyset$; ~~/*initialization*/ \\[.2em]
          $~~$\textbf{ while} there exists $R \in V_{\mathit{inf}} (Q)$ and \textbf{visit}(R) == 0\\[.2em]
          $~~~~~$\textbf{do} $P=\mathit{NF}(R)$;
          $~~\textbf{visit}(R) =1$;\\[.2em]
          $~~~~~\textbf{switch} (P)$\\[.2em]
          ~~~~~~~~~~\textbf{case} $P \equiv \bigvee\limits_{j=1}^h P_{ej} \wedge \emptyy$: \textbf{break};\\[.2em]
          ~~~~~~~~~~\textbf{case} $P \equiv \bigvee\limits_{i=1}^k P_i \wedge \bigcirc P_i'$ or
                              $P \equiv (\bigvee\limits_{j=1}^h P_{ej} \wedge \emptyy) \vee (\bigvee\limits_{i=1}^k P_i \wedge \bigcirc P_i')$:\\[.2em]
          ~~~~~~~~~~~~~~~~~~~~~~~\textbf{foreach} $i~(1 \leq i \leq k)$ \textbf{do}\\[.2em]
          ~~~~~~~~~~~~~~~~~~~~~~~~~~ \textbf{if} $~\neg ( P_i' \equiv \false)~$ and $P_i' \not\in V_{\mathit{inf}}(Q)$ \\[.2em]
          ~~~~~~~~~~~~~~~~~~~~~~~~~~ \textbf{then} $\textbf{visit} (P_i') = 0$;
          \\[.2em]
          ~~~~~~~~~~~~~~~~~~~~~~~~~~ /*$P_i$ is not decomposed to normal form*/ \\[.2em]
          ~~~~~~~~~~~~~~~~~~~~~~~~~~~~~~~~~~  $V_{\mathit{inf}}(Q) = V_{\mathit{inf}}(Q) \cup \bigcup\limits_{i=1}^k \{P_i'\}$; \\[.2em]
          ~~~~~~~~~~~~~~~~~~~~~~~~~~~~~~~~~~  $E_{\mathit{inf}}(Q) = E_{\mathit{inf}}(Q) \cup \bigcup\limits_{i=1}^k \{(R, P_i, P_i')\}$;\\[.2em]
          ~~~~~~~~~~~~~~~~~~~~~~~~~~ \textbf{if} $\neg ( P_i' \equiv \false)~$ and $P_i' \in V_{\mathit{inf}}(Q)$\\[.2em]
          ~~~~~~~~~~~~~~~~~~~~~~~~~~ \textbf{then} $E_{\mathit{inf}}(Q) = E_{\mathit{inf}}(Q) \cup \bigcup\limits_{i=1}^k \{(R, P_i, P_i')\}$;\\[.7em]
          ~~~~~~~~~~~~~~~~~~~~~~~~~~~~~~~~~~~\textbf{when} $P_i' = R$~ \textbf{do}~   /*self-loop*/\\[.5em]
          ~~~~~~~~~~~~~~~~~~~~~~~~~~~~~~~~~~~~~ \textbf{if} $R$ is $Q_1 \;\CHOP\; Q_2$ \textbf{then} $E= E \cup \{R\}$ \textbf{else} $F =F \cup \{R\}$
          \\[.2em]
          ~~~~~~~~~~~~~~~~~~~~~~~~~~~~~~~~~~ \textbf{for} some node $R'' \in V_{\mathit{inf}}(Q)$;\\[.2em]
          ~~~~~~~~~~~~~~~~~~~~~~~~~~~~~~~~~~ \textbf{let}
          $NF(R'')= \bigvee\limits_{j=1}^k R_i \wedge \bigcirc R$ or\\[.2em]
          ~~~~~~~~~~~~~~~~~~~~~~~~~~~~~~~~~~~~~~~
          $NF(R'')=(\bigvee\limits_{j=1}^h R_{ej} \wedge \emptyy) \vee (\bigvee\limits_{i=1}^k R_i \wedge \bigcirc
          R)$;\\[.7em]
          ~~~~~~~~~~~~~~~~~~~~~~~~~~~~~~~~~~ /*nodes $R$ and $R''$ form a loop*/\\[.5em]
          ~~~~~~~~~~~~~~~~~~~~~~~~~~~~~~~~~~~\textbf{when} $P_i' = R'' ~( R'' \neq
          R)$~\textbf{do}
          ~~\\[.2em]
          ~~~~~~~~~~~~~~~~~~~~~~~~~~~~~~~~~~~~~ \textbf{if} $R, R'' \not \in E$  \textbf{then} $F= F \cup \{\{R, R''\}\}$ \\[.2em]
          ~~~~~~~~~~~~~~~~~~~~~~~~~~~~~~~~~~~~~ \textbf{else} $E =E \cup \{\{R,
          R''\}\}$; \\[.2em]
          ~~~~~~~~~ \textbf{break};\\[.2em]
          ~~\textbf{end while}\\[.2em]
          ~~\textbf{return} $G_{\mathit{Rabin}}$;\\[.2em]
          \textbf{end function}\\[.2em]
          \hline
          \end{tabular}
          \end{table*}

           \subsection{The Algorithms}
         To investigate the probabilistic model checking problem for interval-based temporal logics,
         we use Markov chain $M$ as stochastic models and PPTL as a specification language.
         In the following, we present algorithms for the construction and determinization of $\NFG_{\mathit{inf}}$ with Rabin acceptance condition respectively.

         \subsubsection*{Construction of $\NFG_{\mathit{inf}}$ }

          In Table \ref{Alg-NFG-INF},  we present algorithm $\NFG_{\mathit{inf}} (Q)$ for constructing the
          $\NFG_{\mathit{inf}}$ with Rabin acceptance
          condition for any PPTL formula. Algorithm $\mathit{NF}(Q)$ can be found in
          \cite{ZCL07}, which is used for the purpose of transforming formula $Q$ into
          its normal form. For any formula $R \in V_{\mathit{inf}} (Q)$ and $visit(R)=0$, we assume that $P=\mathit{NF}(R)$ is in normal
          form, where $visit(R)=0$ means that formula $R$ has
          not been decomposed into its normal form.
          When $P \equiv \bigvee_{i=1}^{k} P_i \vee \bigcirc P_i'$ or
          $P \equiv (\bigvee_{j=1}^{h} P_{ej} \wedge \emptyy) \vee (\bigvee_{i=1}^{k} P_i \wedge
          \bigcirc P_i')$, if $P_i'$ is a new formula (node), that is, $P_i' \not\in
          V_{\mathit{inf}}$, then by Definition \ref{NFG-inf},  we add the new node $P_i'$ to $V_{\mathit{inf}}$
          and edge $(R, P_i, P_i')$
          to  $E_{\mathit{inf}}$ respectively.
          On the other hand, if $P_i' \in V_{\mathit{inf}}$, then it will
          be a loop. In particular, we need to consider the case of $R \equiv Q_1 ~\CHOP~ Q_2$. Because $Q_1 ~\CHOP~ Q_2$
          ($Q_1$ \emph{chop} $Q_2$, defined in Fig.\ref{fig-formulas}) represents a computation of $Q_1$ followed by $Q_2$,
          and the intervals for $Q_1$ and $Q_2$ share a common state. That is, $Q_1$
          holds from now until some point in future and from that time point
          $Q_2$ holds. Note that $Q_1 ~\CHOP~ Q_2$ used here is a \emph{strong chop} which always
          requires that $Q_1$ be true on some finite subinterval.
          Therefore, infinite models of $Q_1$ can cause $R$ to be false. To solve the problem,
          we employ Rabin acceptance condition to constraint
          that chop formula will not be repeated infinitely many
          times.

          By Theorem \ref{finiteness}, we know that nodes $V(Q)$ is
          finite in $\NFG$. Since $V_{inf}(Q) \subseteq V(Q)$, so
          $V_{\mathit{inf}}(Q)$ is finite as well. This is essential
          since it can guarantee that the algorithm $\NFG_{\mathit{inf}}
          (Q)$ will terminate.

          \begin{Thm}\label{alg-termination}\rm
          Algorithm $\NFG_{\mathit{inf}} (Q)$ always terminates.
          \end{Thm}
          \begin{Proof}
          Let $V_{\mathit{inf}} (Q) =\{v_1,\ldots,
          v_n\}$. When all nodes in $V_{\mathit{inf}}$ are
          transformed
          into  normal form, we have $visit(v_i)==1~(1\leq i\leq
          n)$. Hence, the while loop always terminates.
          \end{Proof}

           We denote the set of infinite paths in an $\NFG_{\mathit{inf}}$ $G$ by
          $\mathit{path}(G) = \{p_1, \ldots, p_m\}$, where $p_i~(1 \leq i \leq m)$ is an
          infinite path from the initial node to some acceptable node in $F$. The following theorem holds.

           \begin{Thm}\label{NFG-equiv}\rm
          $G_{Rabin}$ and $G'_{\mathit{Rabin}}$ are equivalent if and only if
           $\mathit{path}(G_{\mathit{Rabin}})= \mathit{path}(G'_{\mathit{Rabin}})$.
          \end{Thm}

          Let $Q$ be a satisfiable PPTL formula.  By unfolding
          the normal form of $Q$, there is a sequence of formulas $\langle Q, Q_1, Q_1', Q_2, Q_2', \ldots
          \rangle$. Further,
          by algorithm $\NFG_{\mathit{inf}}$, we can obtain an equivalent
          $\NFG_{\mathit{inf}}$
          to the normal form. In fact, an infinite path in
          $\NFG_{\mathit{inf}}$ of $Q$ corresponds to a model of
          $Q$.  We conclude this fact in Theorem
          \ref{NFG-inf-thm}.

          \begin{Thm}\label{NFG-inf-thm}\rm
          A formula $Q$ can be satisfied by infinite models if and
          only if there exists infinite paths in $\NFG_{\mathit{inf}}$ of $Q$ with
          Rabin acceptance condition.
          \end{Thm}

         \subsubsection*{Determinization of $\NFG_{\mathit{inf}}$}

           Buchi automata and $\NFG_{\mathit{inf}}$ both accept $\omega$-words.
           The former is a basis for the automata-theoretic approach for model checking with liner-time temporal logic,
           whereas the latter is
           the basis for the satisfiability and model checking of PPTL formulas.
           Following the thought of the Safra's construction for deterministic Buchi
           automata \cite{Wolfgang02}, we can obtain a deterministic $\NFG_{\mathit{inf}}$ with Rabin acceptance condition
           from the non-deterministic ones.
           However, different from the states in Buchi automata,
           each node in $\NFG_{\mathit{inf}}$ is specified by a formula in PPTL.
           Thus, by eliminating the nodes that contain equivalent formulas,
           we can decrease the number of states in the resulting deterministic $\NFG_{\mathit{inf}}$
           to some degree.

          The construction for deterministic $\NFG_{\mathit{inf}}$ is shown in Table \ref{Alg-DNFG}.
          For any $R\in V_{\mathit{inf}}'(Q)$, $R$ is a Safra tree
          consisting of a set of nodes, and each node $v$ is a set of formulas.
          By Safra's algorithm \cite{Wolfgang02}, we can compute all reachable Safra tree $R'$ that can be reached from $R$ on input
          $P_i$.  To obtain a deterministic $\NFG_{\mathit{inf}}$, we take all pairs $(E_v, F_v)$ as acceptance component, where
          $E_v$ consists of all Safra trees without a node $v$, and $F_v$ all Safra trees with node $v$ marked '!' that denotes
          $v$ will recur infinitely often. Furthermore, we can minimize the number of states in the resulting $\NFG_{\mathit{inf}}$ by
          finding equivalent nodes.
          Let $R = \{v_0, \ldots, v_n\}$ and $R'=\{v_0', \ldots, v_n'\}$ be two Safra's trees, where $R, R' \in V'_{\mathit{inf}}$,
          nodes $v_i= \{Q_1,Q_2, \ldots\}$ and $v_i' =\{Q_1', Q_2', \ldots\}$ be a set of formulas.
          For any nodes $v_i$ and $v_i'$, if we have $v_i = v_i'$, then  the two Safra's trees are the same.
          Moreover,  we have $v_i = v_i'$ if and only if $\bigvee_{j=1}^n Q_j \equiv \bigvee_{j=1}^n Q_j'$.
          The decision procedure for formulas equivalence can be guaranteed by  satisfiability
          theorems presented in \cite{ZCL07}.

           \begin{table}[htpb]
          \centering \caption {Algorithm for Deterministic $\NFG_{\mathit{inf}}$.} \label{Alg-DNFG}
          \begin{tabular}{|l|}
          \hline\\[-.7em]
          \textbf{Function} DNFG(Q)\\
          /*precondition:  $G_{\mathit{Rabin}} = (V_{\mathit{inf}}(Q), E_{\mathit{inf}}(Q), v_0, \Omega)$
          is an  $\NFG_{\mathit{inf}}$ for PPTL formula $Q$.  */ \\
          /*postcondition: DNFG(Q) outputs a deterministic $\NFG_{\mathit{inf}}$ and \\
          ~~~$G_{Rabin}'=(V_{\mathit{inf}}'(Q), E_{\mathit{inf}}'(Q), v_0', \Omega')$ */ \\[.2em]
          \hline\\[-.7em]
          \textbf{begin function} \\[.2em]
          ~~$V_{\mathit{inf}}'(Q) = \{Q\}; E_{\mathit{inf}}'(Q)=\emptyset; v_0' = v_0; E_v = F_v = \emptyset;$  /*initialization*/ \\[1em]
          ~~\textbf{while} $R\in V_{\mathit{inf}}'(Q)$ and there exists an input $P_i$ \textbf{do}\\[.2em]
          ~~~~\textbf{foreach} node $v \in R$ such that  $R \cap F \neq \emptyset$ \\[.2em]
          ~~~~~~\textbf{do} $v'= v \cap F$; $ R'=R \cup \{v'\}$; /* create a new node $v'$ such that $v'$ is a son of $v$*/\\[.2em]
          ~~~~\textbf{foreach} node $v$ in $R'$\\[.2em]
          ~~~~~~\textbf{do} $v = \{P_i' \in V_{inf}(Q)\mid \exists(P, P_i, P_i' )\in E_{inf}(Q), P\in v\}$; /*update $R'$*/\\[.2em]
          ~~~~\textbf{foreach} $v \in R'$ \textbf{do if} $P_i \in v$ such that $P_i \in $ left sibling of $v$ \textbf{then} remove $P_i$ in $v$; \\[.2em]
          ~~~~\textbf{foreach} $v \in R'$ \textbf{do if} $v=\emptyset$ \textbf{then} remove $v$;\\[.2em]
          ~~~~\textbf{foreach} $v\in R'$  \textbf{do if} $u_1,\ldots,u_n$ are all sons of $v$ such that $v=\cup_{i}\{u_i\}~(1\leq i \leq n)$ \\[.2em]
          ~~~~~~~~~~~~~~~~~~~~~~~~~~~~~ \textbf{then} remove $u_i$; mark $v$ with $!$;\\[.2em]
          ~~~~$V_{\mathit{inf}}'(Q) = \{R'\} \cup V_{\mathit{inf}}'(Q)$; $E_{\mathit{inf}}'(Q) = (R, P_i, R')\cup E_{\mathit{inf}}'(Q)$;\\[.2em]
          ~~\textbf{end while}\\[.5em]
          ~~/*Rabin acceptance components*/ \\[.2em]
          ~~ $E_v=\{R \in V_{\mathit{inf}}'(Q) \mid \mbox{R is Safra tree without node $v$}\}$; \\[.2em]
          ~~ $F_v=\{R \in V_{\mathit{inf}}'(Q) \mid \mbox{R is Safra tree with $v$ marked $!$}\}$;\\[.2em]
          ~~\textbf{return} $G_{\mathit{Rabin}}'$;\\
          \textbf{end function}\\[.2em]
          \hline
          \end{tabular}
          \end{table}

          \subsection{Product Markov Chains}


          \begin{Def}\label{product}\rm
          Let $M=(S, \mathit{Prob}, \iota_{init}, AP, L)$ be a Markov chain $M$, and for PPTL formula $Q$,
          $G_{\mathit{Rabin}} = (V_{\mathit{inf}} (Q), E_{\mathit{inf}}(Q), v_0, \Omega)$
          be a deterministic $\NFG_{\mathit{inf}}$, where $\Omega =\{(E_1, F_1), \ldots, (E_k, F_k)\}$.
          The product $M \otimes G_{\mathit{Rabin}}$ is the Markov chain, which is defined as follows.

          \[
             M \otimes G_{\mathit{Rabin}} = (S \times V_{\mathit{inf}}(Q), \mathit{Prob'}, \iota_{init}, \{acc\}, L' )
          \]
          where
          \[
            L'(\langle s, Q'\rangle) = \left\{
                                    \begin{array}{ll}
                                    \{ acc \} & \mbox{ if for some $F_i$}, Q' \in F_i, \\&
                                     \mbox{ and } Q' \not \in E_j \mbox{ for all } E_j,\\ &
                                     ~1 \leq i, j \leq k\\
                                    \emptyset & \mbox{ otherwise}
                                    \end{array}
                                    \right.
          \]

          \[
             \iota_{init}' (\langle s, Q'\rangle) = \left\{
                                                           \begin{array}{ll}
                                                            \iota_{init}~ &\mbox{ if } (Q, L(s), Q') \in E_{\mathit{inf}}\\
                                                            0~&\mbox{ otherwise }
                                                           \end{array}
                                                    \right.
          \]
          and transition probabilities are given by
          \[
          \begin{array}{lrl}
            && \mathit{Prob}'(\langle s', Q'\rangle, \langle s'', Q'' \rangle)\\
             & = &
             \left\{
                    \begin{array}{ll}
                    \mathit{Prob}(s', s'')& \mbox{ if } (Q', L(s''), Q'') \in E_{\mathit{inf}}\\
                    0 & \mbox{ otherwise }
                    \end{array}
             \right.

          \end{array}
          \]

          \end{Def}
          
           A bottom strongly connected components (BSCCs) in $M \otimes G_{\mathit{Rabin}}$ is accepting if it fulfills the
          acceptance condition $\Omega$ in $G_{\mathit{Rabin}}$.

          For some state $s \in M$, we need to compute the probability for the
          set of paths  starting from $s$ in $M$ for which $Q$
          holds, that is, the value of $Pr^M(s \models Q)$. From Definition \ref{product}, it can be reduced to
          computing the probability of accepting runs in the product Markov chain $M \otimes  G_{\mathit{Rabin}}$.

          \begin{Thm}\rm
           Let $M$ be a finite Markov chain, $s$ a state in $M$, $G_{\mathit{Rabin}}$ a deterministic $\NFG_{\mathit{inf}}$ for formula $Q$,
           and let $U$ denote all the accepting BSCCs in $M \otimes G_{\mathit{Rabin}}$. Then, we have
           \[
            Pr^M (s \models G_{\mathit{Rabin}})= Pr^{M\otimes G_{\mathit{Rabin}}} (\langle s, Q' \rangle \models \Diamond U)
           \]
           where $(Q, L(s), Q') \in E_{\mathit{inf}}$.
          \end{Thm}

           \begin{Cor}\rm
           All the $\omega$-regular properties specified by PPTL are measurable.
           \end{Cor}

          \begin{Expl}\rm
          We now consider the example in Figure \ref{eg-chop}. Let $M$ denote Markov chain in Figure \ref{eg-chop}(b).
          The probability that 
           \emph{sequential property} $p ~\CHOP ~q$ holds in Markov chain $M$ can be computed as follows.
           
          First, by the two algorithms above, deterministic $\NFG_{\mathit{inf}}$ with Rabin condition for $p ~\CHOP ~q$ is 
          constructed as in Figure \ref{eg-chop}(a), where the Rabin acceptance condition is $\Omega ={(v_1, v_2)}$.
          Further, the product of the Markov chain and $\NFG_{\mathit{inf}}$
          for formula $p ~\CHOP ~q$ 
          is given in Figure \ref{product2}.

          \begin{figure}[htpb]
          \centering\includegraphics[width=8cm]{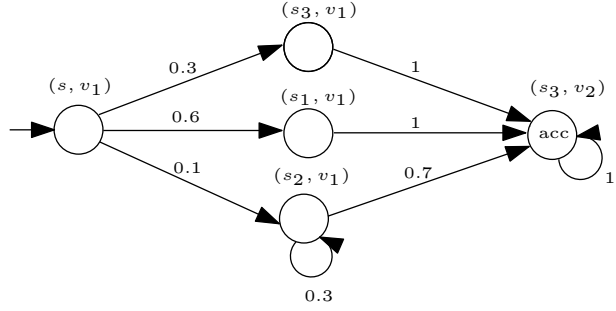} \caption{The Product of Markov chain and $\NFG_{\mathit{inf}}$ in Figure \ref{eg-chop}.}
          \label{product2}
          \end{figure}

          From Figure \ref{product2}, we can see that
           state $(s_3, v_2)$  is the unique accepting BSCC. Therefore,  we have
          \[
          \begin{array}{lrl}
          &&Pr^M (s \models G_{\mathit{Rabin}})\\
          &=&Pr^{M \otimes G_{\mathit{Rabin}}} ((s, v_1) \models \Diamond (s_3, v_3)) \\
          &=&1
          \end{array}
          \]
          That is, sequential property $p~\CHOP~q$ is satisfied almost surely by the Markov chain $M$ in Figure \ref{eg-chop}(b).
   \end{Expl}

          \section{Conclusions}
           This paper presents an approach for probabilistic model
           checking based on PPTL. Both propositional LTL and PPTL can specify linear-time properties.
           However, unlike probabilistic model checking on propositional LTL,
           our approach uses NFGs, not Buchi automata, to characterize models of
           logic formulas.  NFGs possess some merits that are
           more suitable to be employed in model checking for interval-based temporal logics.

           Recently, some promising formal verification techniques
           based on NFGs have been developed, such as  \cite{CD10,ZN08}.
           In the near future, we will extend the existing model checker for PPTL with probability,
           and according to the algorithms proposed in this paper,
           to verify the regular safety properties in probabilistic systems.

         \section*{Acknowledgement}
          Thanks to Huimin Lin and Joost-Pieter Katoen for their helpful suggestions.


\begin{thebibliography}{99}

\bibitem{Hansson94} H. Hansson and B. Jonsson. (1994), A Logic for Reasoning about Time and Reliability.
Formal Aspects of Computing.
Vol. 6, pages 102-111.

\bibitem{Aziz00} A. Aziz, K. Sanwal, V. Singhal and R. K. Brayton. (2000), Model Checking Continous Time Markov Chains.
ACM Trans. Comput. Log. Vol. 1(1): 162-170.

\bibitem{Mos83} B. Moszkowski. (1983), Reasoning about digitial circuits. PhD Thesis. Stanford University.
TRSTAN-CS-83-970.


\bibitem{Mos86} B.Moszkowski. (1986),
Executing temporal logic programs.   Cambridge University Press.

\bibitem{PRISM} M. Z. Kwiatkowska, G. Norman and D. Parker. (2004), Probabilistic Symbolic Model Checking with PRISM:
a hybrid approach. \emph{STTT}. Vol. 6(2): 128-142.

\bibitem{Katoen} C. Baier, J. P. Katoen. (2008), Principles of Model Checking.
The MIT Press.



\bibitem{MRMC} J. -P. Katoen, M. Khattri and I. S. Zapreev. (2005), A Markov Reward Model Checker. In \emph{QEST}, pages 243-244.

\bibitem{KS60} J. G. Kemeny and J. L. Snell. (1960), Finite Markov Chains. Van Nostrad, Princeton.



\bibitem{wolper}
P. L. Wolper. (1983), Temporal logic can be more expressive.
\emph{Information and Control}, vol.56, pages 72-99. 1983.




\bibitem{DYK08} Z.Duan, X.Yang and M.Koutny. (2008), Framed Temporal Logic
Programming. {\em Science of Computer Programming}, Volume 70(1),
pages 31-61, Elsevier North-Holland.

\bibitem{YD08}
X. Yang and Z. Duan. (2008) Operational Semantics of Framed Tempura. {\em
Journal of Logic and Algebraic Programming}, Vol.78(1):22-51,
Elsevier North-Holland.


\bibitem{ZCL07}
Z. Duan, C. Tian, L. Zhang. (2008), A Decision Procedure for Propositional
Projection Temporal Logic with Infinite Models. \emph{Acta
Informatic}, Springer-Verlag, 45, 43-78.

\bibitem{CD10}
C.Tian and Z. Duan. (2010), Making Abstraction Refinement Efficient in Model
Checking CoRR abs/1007.3569.


\bibitem{ZN08} C. Tian, Z. Duan. (2007),
 Model Checking Propositional Projection Temporal Logic Based on SPIN. ICFEM
 pages: 246-265.


\bibitem{Wolfgang02}
E. Gr$\ddot{a}$del, W. Thomas, T. Wilke (Eds.). (2002), Automata Logics, and
Infinite Games: A Guide to Current Research. Springer-Verlag Berlin
Heidelberg.

\end{thebibliography}
 \end{document}